\documentstyle[12pt]{article}

\font\twelve=cmbx10 at 15pt
\font\ten=cmbx10 at 12pt

\def\cit#1{$^{[#1]}$}

\def\pr{Phys.\ Rev.\ }
\def\prl{Phys.\ Rev.\ Lett.\ }
\def\pl{Phys.\ Lett.\ }
\def\etal{{\it et al}.}

\newcommand{\beq}[1]{\begin{equation}\label{#1}}
\newcommand{\eeq}{\end{equation}}

\newcommand{\rf}[1]{(\ref{#1})}

\begin{document}

\begin{titlepage}

\begin{center}

\renewcommand{\thefootnote}{\fnsymbol{footnote}}

{\ten Centre de Physique Th\'eorique\footnote{
Unit\'e Propre de Recherche 7061} - CNRS - Luminy, Case 907}
{\ten F-13288 Marseille Cedex 9 - France }

\vspace{1 cm}

{\twelve EXPANDING PROTONS SEEN BY HERA}

\vspace{0.3 cm}

\setcounter{footnote}{0}
\renewcommand{\thefootnote}{\arabic{footnote}}

{\bf Claude BOURRELY, Jacques SOFFER and Tai Tsun WU\footnote{
CERN - Geneva, Switzerland and Gordon McKay Laboratory, Harvard
University, Cambridge, MA 02138, U.S.A.}$^,$\footnote{
Work supported in part by the US Department of Energy under Grant
N$^{\circ}$. DE-FG02-84ER40158 and the Joint Service Electronics
Program under Grant N00014-89-J-1023.}
}

\end{center}

\vspace{1 cm}

\setlength{\baselineskip}{24pt}

\centerline{\bf Abstract}

We show that the rising total photoproduction cross section recently
observed at HERA is entirely consistant with the impact-picture in
the theory of expanding protons which was proposed more than twenty
years ago. More accurate data, which is expected to be forthcoming
soon, will give this prediction a stringent test.

\setlength{\baselineskip}{15pt}

\vspace{1,5 cm}


\bigskip

\noindent Number of figures : 2

\bigskip

\noindent August 1994

\noindent CPT-94/P.3060

\noindent CERN-TH/7387/94

\bigskip

\noindent anonymous ftp or gopher: cpt.univ-mrs.fr

\end{titlepage}

\setlength{\baselineskip}{24pt}

Recently at HERA, the $H 1$ and ZEUS detectors have measured the
photo\-production total cross section at an average $\gamma p$ center
of mass energy of $180\ G e V$\cit{1,2}. They confirm the previous
and less accurate result of ZEUS\cit{3} and indicate that
this total cross section is rising. This rise has caused a lot of
excitement and vastly different rates of increase have been
proposed\cit{4-9} which are based either on minijet contributions or
on Regge - type analyses.

In this letter we will present a very simple approach to this
increasing $\gamma p$ total cross section. Increases of
hadron-hadron total cross sections have been theoreticaly predicted
some years ago\cit{10,11,12} and these predictions have been
accurately verified by experiment\cit{13}. The theoretical
predictions were based on the multi-tower diagrams shown in fig.1a
for $pp$ and $\overline p p$ and fig.1b for $\pi^{\pm} p$ elastic
scattering. The multi-tower diagrams for $\gamma p$ are shown in
fig.1c. The similarity of the diagrams in fig.1b and fig.1c implies
that the $\gamma p$ total cross section can be related approximately
to those of $\pi^{\pm} p$. The major differences between these two
figures are

\begin{itemize}

\item[i)] The electromagnetic coupling $e$ appears twice in fig.1c.

\item[ii)] The photon is its own antiparticle.

\end{itemize}

Because of those two differences it is natural to expect
\beq{1}
\sigma_{\hbox{tot}} (\gamma p) = ( \hbox{const.} ) \alpha \left(
\sigma_{\hbox{tot}} (\pi^+ p) + \sigma_{\hbox{tot}} (\pi^- p)
\right)
\eeq
where $\alpha$ is the fine structure constant. This eq.(1) should hold for all
center
of mass energies above about
$10GeV$. The constant is to be determined from the low energy $\gamma p$ total
cross section data\cit{14}.
 Note that
the contribution from $\rho$-exchange is cancelled in the sum, as it
should be. For $\sigma_{\hbox{tot}} (\pi^{\pm} p)$ we have used the
early impact picture predictions\cit{15} where a simple $s$
dependence $s^{0.08}$ was first given and later extensively used by
various authors. Using these results it is found that the constant
in eq.\rf{1} is $1/3$ and thus the predicted $\gamma p$ total
cross section is shown in fig.2. The HERA data is surely going to
improve rapidly and we are eager to see a much more accurate
determination of the $\gamma p$ cross section.

Thus we have encountered the expanding proton unambigously in two
different experimental situations. In both cases, $pp$ and $\gamma
p$, the next major increase in energy will come from LHC. In these
cases the center of mass energy will reach $16\ T e V$ for $pp$ are
more than $1\ T e V$ for $\gamma p$ using electrons from LEP.
Moreover using extracted $\pi^+$ and $\pi^-$ beams from LHC, the $\pi
p$ cross sections could be measured up to about $\sqrt{s} = 120\ G e
V$. This will provide a most stringent test of the accuracy of our
relation~\rf{1}.

\section*{Acknowledgments}

Tai~Tsun~Wu wishes to thank Arthur~Jordan for a helpful discussion
and is grateful to John Ellis, Maurice Jacob, Andr\'e Martin and many
other members of the Theory Division for their hospitality at
CERN.

\vskip1cm

\section*{Figure Captions}

\begin{itemize}

\item[Fig.1] Multi-tower diagrams for (a) $pp$ and $\overline p p$,
(b) $\pi^{\pm} p$ and (c) $\gamma p$ scattering.

\item[Fig.2] $\gamma p$ total cross section as a function of the
center of mass energy. The data are from refs.~\cite{1}, \cite{2} and
\cite{14} and the curve is from eq.\rf{1}.

\end{itemize}


\newpage


\begin{thebibliography}{99}

\bibitem{1} $H 1$ Collaboration, T.~Ahmed \etal, \pl {\bf B299} (1993)
374.

\bibitem{2} ZEUS Collaboration, M.~Derrick \etal, Preprint
DESY~94-032 (march 1994).

\bibitem{3} ZEUS Collaboration, M.~Derrick \etal, \pl {\bf B293}
(1992) 465.

\bibitem{4} M.~Drees and F.~Halzen, \prl {\bf 61}
(1988) 275.

\bibitem{5} R.~Gandhi and I.~Sarcevic, \pr {\bf D44}
(1991) 10R.

\bibitem{6} J.R.~Forshaw and J.K.~Storrow, \pl {\bf B268}
(1991) 116.

\bibitem{7} R.S.~Fletcher, T.K.~Gaisser and F.~Halzen, \pr {\bf
D45} (1992) 377.

\bibitem{8} H.~Abramowicz, E.M.~Levin, A.~Levy and U.~Maor, \pl
{\bf B269} (1991) 465.

\bibitem{9} A.~Donnachie and P.V.~Landshoff, \pl {\bf B296}
(1992) 227.

\bibitem{10} H.~Cheng and T.T.~Wu, \prl {\bf 24}
(1970) 1456.

\bibitem{11} C.~Bourrely, J.~Soffer and T.T.~Wu, \prl {\bf
54} (1985) 757.

\bibitem{12} H.~Cheng and T.T.~Wu, Expanding protons~: Scattering at
high energies (MIT Press, Cambridge, MA, 1987).

\bibitem{13} See for example N.A.~Amos \etal, \pl {\bf
B243} (1990) 158.

\bibitem{14} D.O.~Caldwell \etal, \prl {\bf
40} (1978) 1222.

\bibitem{15} H.~Cheng, J.K.~Walker and T.T.~Wu, \pl {\bf B44}
(1973) 97.

\end{thebibliography}
\end{document}